# Multiple Genome Analytics Framework: The Case of All SARS-CoV-2 Complete Variants

Konstantinos F. Xylogiannopoulos
*Department of Computer Science*
*University of Calgary*
Calgary, AB, Canada
kostasfx@yahoo.gr
0000-0003-2376-898X

***Abstract* — Pattern detection and string matching are fundamental problems in computer science and the accelerated expansion of bioinformatics and computational biology have made them a core topic for both disciplines. The SARS-CoV-2 pandemic has made such problems more demanding with hundreds or thousands of new genome variants discovered every week, because of constant mutations, and there is a desperate need for fast and accurate analyses. The requirement for computational tools for genomic analyses, such as sequence alignment, is very important, although, in most cases the resources and computational power required are enormous. The presented Multiple Genome Analytics Framework combines data structures and algorithms, specifically built for text mining and pattern detection, that can help to efficiently address several computational biology and bioinformatics problems concurrently with minimal resources. A single execution of advanced algorithms, with space and time complexity $O(n \log n)$, is enough to acquire knowledge on all repeated patterns that exist in multiple genome sequences and this information can be used from other meta-algorithms for further meta-analyses. The potential of the proposed framework is demonstrated with the analysis of more than 300,000 SARS-CoV-2 genome sequences and the detection of all repeated patterns with length up to 60 nucleotides in these sequences. These results have been used to provide answers to questions such as common patterns among all variants, sequence alignment, palindromes and tandem repeats detection, different organism genome comparisons, polymerase chain reaction primers detection, etc.**

## I. Introduction

The COVID-19 pandemic has highlighted governmental, scientific, economic and political focus on the biotechnology industry and its efforts to address the virus consequences as soon as possible. Major pharmaceutical and biotechnology companies worldwide have invested huge amounts in new technologies for the past couple of decades and the first promising results from technologies such the mRNA vaccines have become visible. Indeed, the fast expansion of the biotechnology industry with the help of advanced computing infrastructures, such as cloud computing, has opened a new era in the domain.

Some of the most common problems addressed in computer science over time are related to pattern matching and searching. In bioinformatics, there has been a plethora of completely diverse methodologies and algorithms since early 1970, which were developed to deal with the simplest problems, such as to determine if a specific string exists in a biological sequence, to more complex problems such as the multiple sequence alignment. Furthermore, the development of artificial intelligence and deep learning provides more sophisticated tools for image analysis or clinical data analytics.

The analyses of biological sequences such as DNA, RNA, proteins, etc. are considered standard string problems in computer science since such sequences are built from predefined discrete alphabets (the nucleotides or the amino-acids encoding). What make these string problems challenging in bioinformatics and computational biology, from a computer science perspective, is the size of the strings and the computationally intensive procedures to solve them. Moreover, in most cases solutions cannot be provided in short time with regular computational resources. For example, the complete, combined, human genome, a 3.1Gbp long string, was initially sequenced in 2001 [1] and it was practically impossible to be analyzed by desktop computers as a single piece of information since only supercomputers could store and process data structures of such long strings in memory. For example, the construction of a suffix tree data structure for the first human chromosome with an approximate size of 250Mbp, requires 26GB of memory [4]. Despite the introduction of 64-bit processor architecture at that time, 64-bit operating systems that could handle more than 4GB RAM were introduced a few years later. Nowadays, advanced hardware and clustering framework systems are used for such big data analyses. New technologies such as Next Generation Sequencing (NGS) from leading companies require advanced computational tools and algorithms, specifically designed for string matching problems in order to perform sequence alignment in multiple (usually millions) genomic fragments simultaneously.

The analysis of the full, single, human genome with the detection of all repeated patterns was presented for the first time in 2019 [31]. However, that initial attempt was just a proof of concept and technology for a specific algorithm. The currently presented Multiple Genome Analytics (MuGA) Framework will show that it is possible, with limited resources and in short time, to analyze hundreds of thousands of complete genomes and detect all repeated patterns that exist in them. Moreover, how the combination of an advanced data structure and the results of such analysis can help to solve many pattern detection problems will be presented. Finally, the potential of the tools described in specific types of string problems with applications on a large dataset comprised from all SARS-CoV-2 full genome variants will also be presented.

In order to achieve such results, the Multiple Genome Analytics Framework initially uses the Multivariate Longest Expected Repeated Pattern Reduced Suffix Array (LERP-RSA) data structure in combination with the All Repeated Patterns Detection (ARPaD) algorithm [26], [27], [28]. In brief, LERP-RSA is a variation of the standard Suffix Array [25] data structure using the actual, lexicographically sorted, suffix strings. The ARPaD algorithm, both in its recursive and non-recursive variant, has the ability to scan the LERP-RSA only once and detect every pattern that occurs at least twice in it. Additionally, the algorithm is pattern agnostic, i.e., it does not require an input parameter, rather it scans the data structure once and returns all results in a deterministic way regardless of string or pattern attributes, e.g., frequency, length, alphabet, overlapping or not, etc.

So far, LERP-RSA and ARPaD have been extensively used in many, diversified, domains with vast datasets in most cases and exceptional results, regardless of the hardware limitations, making them a state-of-the-art approach for big data problems in text mining and pattern detection [28]. An example of such a problem is the analysis in 2016 of a single, continuous, string of one trillion characters, constructed from the first digits of $\pi$, which is 4,000 times larger than the largest human chromosome [28]. Such an analysis is unique in literature and the results were validated three years later by the Google pi-api using the Google Cloud Platform [32].

The contribution of the current work is to introduce an innovative framework that can be used to address as many string problems as possible simultaneously. As an example, which also falls into the category of big data analytics, firstly the analysis of 302,373 SARS-CoV-2 genome variants has been executed to discover all repeated patterns. Subsequently, these results have been used by meta-algorithms for additional meta-analytics, such as:

a) discovery of the longest patterns, which exist among every variant of SARS-CoV-2,

b) comparisons among different organisms such as MERS, A-CoV, A-Influenza, HRSV and Human,

c) identification of every frequent and infrequent pattern,

d) detection of restriction enzyme-associated loci,

e) descriptive statistics for mutations and sequence alignment,

f) the detection of special patterns such as:

    i. palindromes,

    ii. tandem repeats,

    iii. polymerase chain reaction (PCR) primers.

Despite any possible limitations of the proposed framework, the benefit of using it on many, diverse, problems concurrently can overcome any initial hesitation.

It is important to mention that it is difficult to directly compare the presented Multiple Genome Analytics Framework with other approaches from an absolute execution time perspective. As far as it is known in literature, ARPaD algorithm is unique in discovering all repeated patterns and it cannot be compared directly to any other algorithm used in computer science and bioinformatics while LERP-RSA is a special data structure designed for the ARPaD algorithm family of variants and additional algorithms. There are several reasons why this claim is justified as will be presented later.

Additionally, the MuGA Framework introduces several innovations, such as:

a) the ARPaD algorithm is highly optimized since it has log-linear time complexity $O(n \log n)$ and also requires log-linear space because of the LERP-RSA $O(n \log n)$ space complexity,

b) the LERP-RSA data structure needs to be constructed only once since it can be stored on off-line media for future use while the ARPaD algorithm also needs to be executed only once on the LERP-RSA,

c) the ARPaD algorithm is deterministic in finding patterns, i.e., it always returns the same results and, therefore, if a pattern exists at least twice it will be found and if a pattern is not in the ARPaD results then it does not exist at least twice in the dataset,

d) current trend of Machine Learning algorithms rely on training and simulation datasets, while the ARPaD algorithm does not require any training or preprocessing,

e) because of the ARPaD algorithm deterministic nature, any commonly used metrics of data analytics such as accuracy, sensitivity, specificity, precision, $F_1$ score, etc. are perfect (100%) in their evaluation since all, correct, repeated patterns have been discovered without any faulty result and there is no need to use simulated data or perform cross-validation techniques to test its accuracy and efficacy,

f) the LERP-RSA data structure and the ARPaD algorithm results can be stored on local and/or remote off-line media and/or databases and they can be used at will and in full parallel mode from other algorithms or query executers,

g) the meta-analytics algorithms introduced in the current work use the ARPaD results as input, which is also a direct advantage in comparison to other algorithms and methodologies because of the revealed knowledge by the repeated patterns,

h) the meta-analytics algorithms have self-parallelization execution attributes and, furthermore, they can be used in parallel to solve many diverse bioinformatics and computational biology problems because they share the same data structure.

The aforementioned innovations of the MuGA framework are not just the key attributes that significantly differentiate the framework from standalone algorithms and processes giving it a competitive advantage in the field of big data analytics, but they are also some of the reasons that justify the difficulty of direct comparison to other approaches, as will be further explained in the next sections.

The rest of the paper is organized as follows: Section II presents related work in string matching. Section III defines the problem and gives the motivation behind it. Section IV presents the data structures and algorithms for pattern detection in biological sequences which form the proposed framework and solve specific problems. Section V presents several applications conducted on the available dataset of all, complete, SARS-CoV-2 variants and discusses the results per problem application. Finally, Section VI presents the conclusions and future extensions of the presented work.

## II. Related Work

In bioinformatics the use of computers to perform analyses of biological sequence, more particular address string matching problems, always had a crucial role. Many new algorithms and methodologies are presented every year that improve older approaches or introduce new [1], [3], [4]. Mainly, these methods and algorithms can be classified into two broad categories, the exact matching and the approximate matching [1], [4]. The first category is related to string problems where we seek to find patterns matching entirely the input string such as, for example, specific sequence matching a protein transcription promoter. The second category can be much more complicated since many mutations, insertions, deletions and base changes may have occur making exact matching difficult, yet, very important, for example, to detect codon sequences which can produce the same protein. However, no algorithm is widely known that can perform a generic, single step, detection of all repeated patterns.

More precisely, exact matching algorithms have dominated the field since early '70s. Many different approaches have been developed such as character or index based. This kind of methodologies include brute force algorithms where characters of the matching pattern are directly compared to the reference sequence. This leads to heavy computational algorithms, mainly because of the absence of any preprocessing and special data structures. The standards for such algorithms are the Boyer-Moore algorithm, usually used as a benchmark for efficiency measurement, that uses a shifting step based on a table holding information about mismatch occurrences and the Knuth-Morris-Pratt algorithm that uses a supplementary table to record temporal information during execution [1], [3], [4], [5], [6]. Another algorithm, variation of the first one mentioned, is the Boyer-Moore-Smith [7] while another extension is the Apostolico-Giancarlo algorithm based on both of the BM and KMP algorithms [8]. Additionally, we have the Raita algorithm based on dependencies that occur among successive characters [9]. More recent algorithms are the BBQ algorithm which introduces parallel pointers that perform searching from opposite directions [10] and several hybrid methods such as the KMPBS [11] and Cao et al. using statistical inference [12].

Except the brute force algorithms we have another important category, the hashed based [1], [3], [4]. Such algorithms are based on the hashing concept in order to produce hashing values and compare patterns rather than performing a direct character comparison. The main benefit from such approach is the considerable improvement of calculation time [13], yet, as with most hashing algorithms, they suffer from the hashing collision problem. Typical examples of such algorithms is the Karp-Rabin which is based on modular arithmetic to perform hashing [14] and the Lecroq algorithm, which first splits the sequence to subsequences and then the pattern matching is performed on each sequence [15]. Classic algorithms are also the non q-gram algorithms such as the Wu and Manber [16] where the searching pattern is completely encoded for pattern matching purposes. Furthermore, more recently developed algorithms are the multi-window integer comparison algorithm based on suffix strings data structures such as the Franek-Jennings-Smyth string matching algorithm [18] and the automata skipping algorithm developed by Masaki et al. [17]. More advanced hybrid approaches have also been presented that combine best practices from different approaches in order to optimize their performance such as, for example, Navarro's algorithm [19] which can bypass characters using suffix.

A very well-known and heavily used algorithm is implemented and used by the National Center for Biotechnology Information (NCBI). The Basic Local Alignment Search Tool (BLAST) and its variants [21] is used for comparing basic sequences, such as nucleotides sequences, found in DNA and/or RNA. The algorithm takes as inputs the desired string to search and the sequence to search into. Additionally, BLAST can execute inexact string matching, something usually extremely computationally intensive, for multiple sequence alignment purposes. Another algorithm, more accurate than BLAST, yet, more resources hungry and slower, is the Smith-Waterman algorithm [20]. Several variations of BLAST also exist, such as the SmartBLAST that it can be used for protein matching and Primer-BLAST that it can be used for primers specific to PCR templates [33].

An important aspect of pattern detection is the discovery of specific type of patterns in biological sequences such as palindromes and tandem repeats. The importance of such discoveries can be presented with one of the latest marbles in biology, the discovery of the clustered regularly interspaced short palindromic repeats (CRISPR) in bacteria and the use of CRISPR-Cas9 protein that allows to interfere with DNA in a molecular level [22]. However, in the case of CRISPR problem it is necessary to identify only palindromes that their length is in between a specific range and they repeat with a relative periodicity. The detection of single occurred, very short or very long palindromes is not important.

Another well studied problem is the detection of short tandem repeats, something very difficult over a whole genome. This kind of repeats are classic examples of repeats in protein encoding regions and are closely related to serious diseases, such as the Huntington's disease [23]. An example of methods for tandems detection can be found in [23] which is based on DNA alignment using LAST software.

## III. Problem Definition

So far, we have presented several algorithms that are used in bioinformatics and computational biology. Yet, all these algorithms have as a common attribute the input pattern that is under investigation. Such type of algorithms can address specific problems and require each time to access the full dataset of one or more sequences to operate and produce results, which could be inefficient.

To address bioinformatics and computational biology problems, it would be more preferable to have a data structure or a database of information that can be used for as many queries as possible and be transformed to valuable knowledge. Moreover, the full process should be able to:

a) be contacted on commodity computers with limited resources

b) keep the cost low

c) allow scale up to deal with larger datasets without the need for new hardware resources

d) address several different computational biology and bioinformatics problems concurrently

## IV. Proposed Framework

The framework that will be introduced in the next sections, is built on the foundation of the Longest Expected Repeated Pattern Reduced Suffix Array (LERP-RSA) data structure [26], [27], [28] and the related family of algorithms such as ARPaD, SPaD and MPaD that are specifically designed for the LERP-RSA [27], [28]. Several applications of the aforementioned data structure and algorithms will be presented, as a pipeline of execution, that can either extract useful information directly from the dataset or the results generated, or can be used as an input for other algorithms for several type of meta-analytics in biological sequences.

### A. LERP-RSA Data Structure

The Longest Expected Repeated Pattern Reduced Suffix Array (LERP-RSA) is a special purpose data structure for pattern detection, which has been developed and optimized to work with a variety of algorithms. Manber and Myers [25] defined the suffix array of a string as the array of the indexes of the lexicographically sorted suffix strings, which allows to perform several tasks on the string, such as pattern matching. The LERP-RSA is a variation of the suffix array, yet, it uses the actual suffix strings and not only the position indexes. The quadratic space complexity of the data structure, with regard to the input string, can be reduced to log-linear with the use of the LERP reduction, derived from the Probabilistic Existence of Longest Expected Repeated Pattern Theorem [27], [28]:

**Theorem:** Let $S$ be a string of size $n$ constructed from a finite alphabet $\Sigma$ of size $m \geq 2$ and let $s$ be a substring (of $S$) of length $l$. Let $X$ be the event that *"substring $s$ occurs at least twice in string $S$."* If $S$ is considerably long and random, and $s$ is reasonably long then the probability $P(X)$ of the event $X$ is extremely small.

The theorem builds us the necessary foundation to calculate the longest expected repeated pattern given a very small probability that a repeated pattern exists with longer length. Therefore, the length of the suffix strings used to create the LERP-RSA, can be reduced significantly by using the following, briefly stated, Lemma [27], [28]:

**Lemma:** Let $S$ be a random string of size $n$, constructed from a finite alphabet $\Sigma$ of size $m \geq 2$, and an upper bound of the probability $P(X)$ is $\overline{P(X)}$, where $X$ the event "LERP is the longest pattern that occurs at least twice in $S$." An upper bound for the length $l$ of the Longest Expected Repeated Pattern (LERP) length we can have with probability $P(X)$ is:

$$\bar{l} = \overline{LERP} = \left\lceil log_m \frac{n^2}{2\overline{P(X)}} \right\rceil$$

where $l \ll n$ and $\overline{P(X)} > 0$.

The abovementioned Lemma is directly inducted from the Theorem and it has been proven in [27], [28]. Of course, the calculation of the longest repeated pattern can be performed by other methods, however, the use of the Lemma has some advantages since, e.g., building the suffix tree and determining the longest repeated pattern of a string on the suffix tree is a heavy computational process and in most of the cases it is impossible because of the string size. For example, in [28] the longest repeated patterns that exist in the first one trillion digits of π have been calculated, knowing in advance the upper limit for their length, while any other algorithmic approach is beyond any possibility with the currently available hardware. Yet, the Theorem and the Lemma have as a prerequisite that the string is random which limits the application for strings that do not have a random behavior. Briefly described, randomness means that every character of the alphabet occurs with the same frequency and this property should be valid for reasonably long substrings, following the normality of irrational numbers property as presented by the Calude's Theorem [24]. Although this is true for most of the cases, unfortunately, biological sequences do not have random behavior and this problem can be solved with the MLERP process as it is described in [26], [28] and it has been used to analyze the full human genome [31].

| | | | | | |
|---|---|---|---|---|---|
| 0 | actactggtgt | 0 | actact | 0 | act |
| 1 | ctactggtgt | 1 | ctactg | 1 | cta |
| 2 | tactggtgt | 2 | tactgg | 2 | tac |
| 3 | actggtgt | 3 | actggt | 3 | act |
| 4 | ctggtgt | 4 | ctggtg | 4 | ctg |
| 5 | tggtgt | 5 | tggtgt | 5 | tgg |
| 6 | ggtgt | 6 | ggtgt | 6 | ggt |
| 7 | gtgt | 7 | gtgt | 7 | gtg |
| 8 | tgt | 8 | tgt | 8 | tgt |
| 9 | gt | 9 | gt | 9 | gt |
| 10 | t | 10 | t | 10 | t |
| | (a) | | (b) | | (c) |

**Fig. 1** Full Suffix Array (a), Arbitrary Reduced Suffix Array (b) and LERP Reduced Suffix Array (c) for *actactggtgt*

The process of constructing the LERP-RSA with the use of the Lemma can be described with the following example. Let's assume that the input string is *actactggtgt*. If we construct the array of the suffix strings then we will receive the structure of Fig.1.a where all suffix strings have been recorded, without sorting. Obviously, this structure has a quadratic space complexity of exact size $O(n(n+1)/2)$ or $O(n^2)$. If the size of the string becomes medium size, e.g., 10,000 characters, as an average human gene, then the space needed just to store the suffix strings, without sorting them, explodes to 100 million. What we can do to bypass the problem, for the initial example, is to reduce the size of the suffix strings to an arbitrary size to, e.g., five characters and create the structure of Fig.1.b. However, in this case we have the following to consider: (a) if the repeated patterns that exist and we want to discover are longer then we will miss all of them with length longer than the five characters and (b) if the repeated patterns are shorter then we are wasting space and time for sorting and analysis. This can be solved with the Lemma and the construction of the LERP-RSA of Fig.1.c, since if we reduce the size of suffix strings to three characters, for example, then the longest pattern that exists and is the *act* can be located at position 0 and 3. The use of value three is an example to illustrate the use the Lemma since it is not accurate for the specific, very sort, example. Since the LERP has length $O(\log n)$, with regard to the size of the input string, then the space complexity of the entire LERP-RSA is $O(n \log n)$.

The LERP-RSA data structure has some unique features that allows to be characterized as a state-of-the-art data

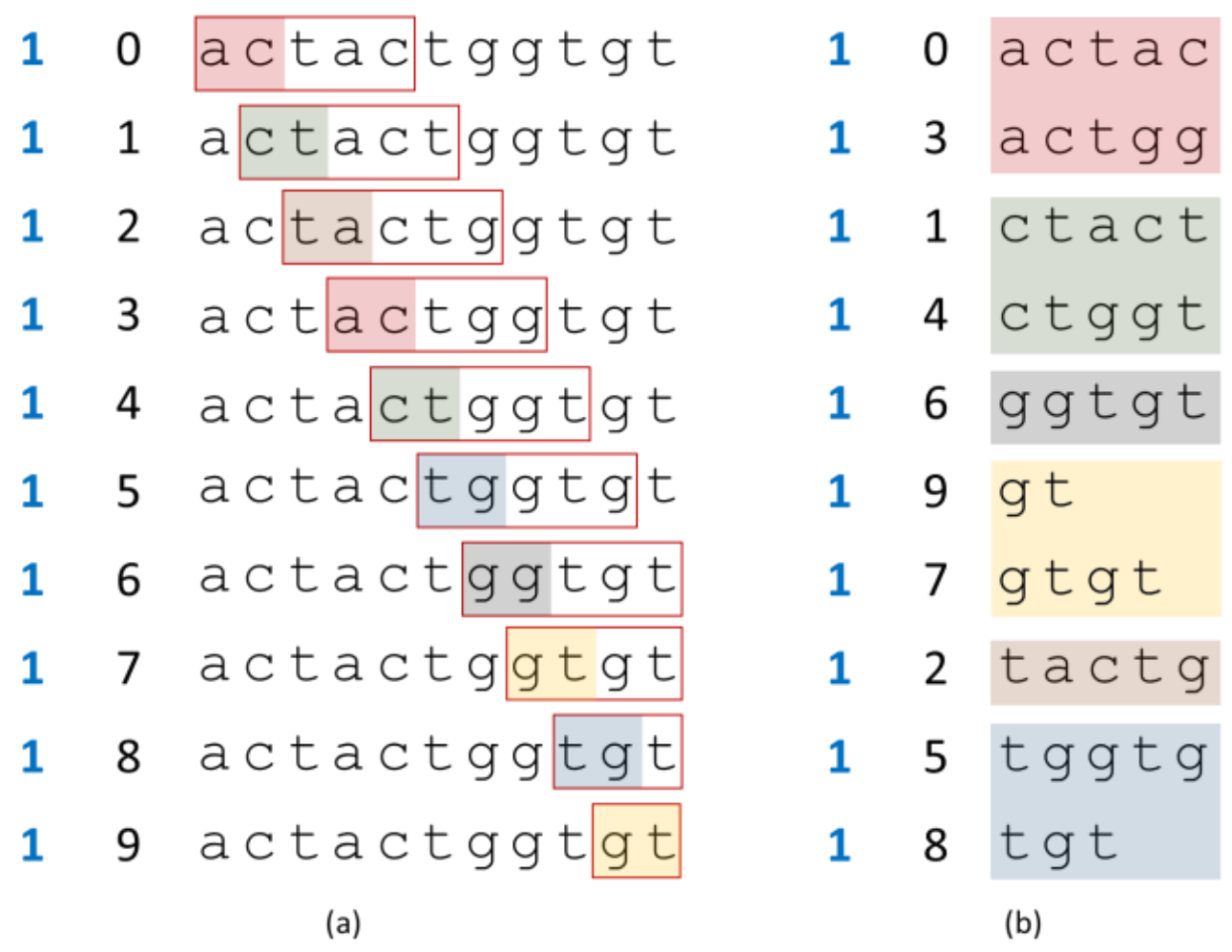

**Fig. 2** LERP-RSA construction for *actactggtgt* with Classification Level 2

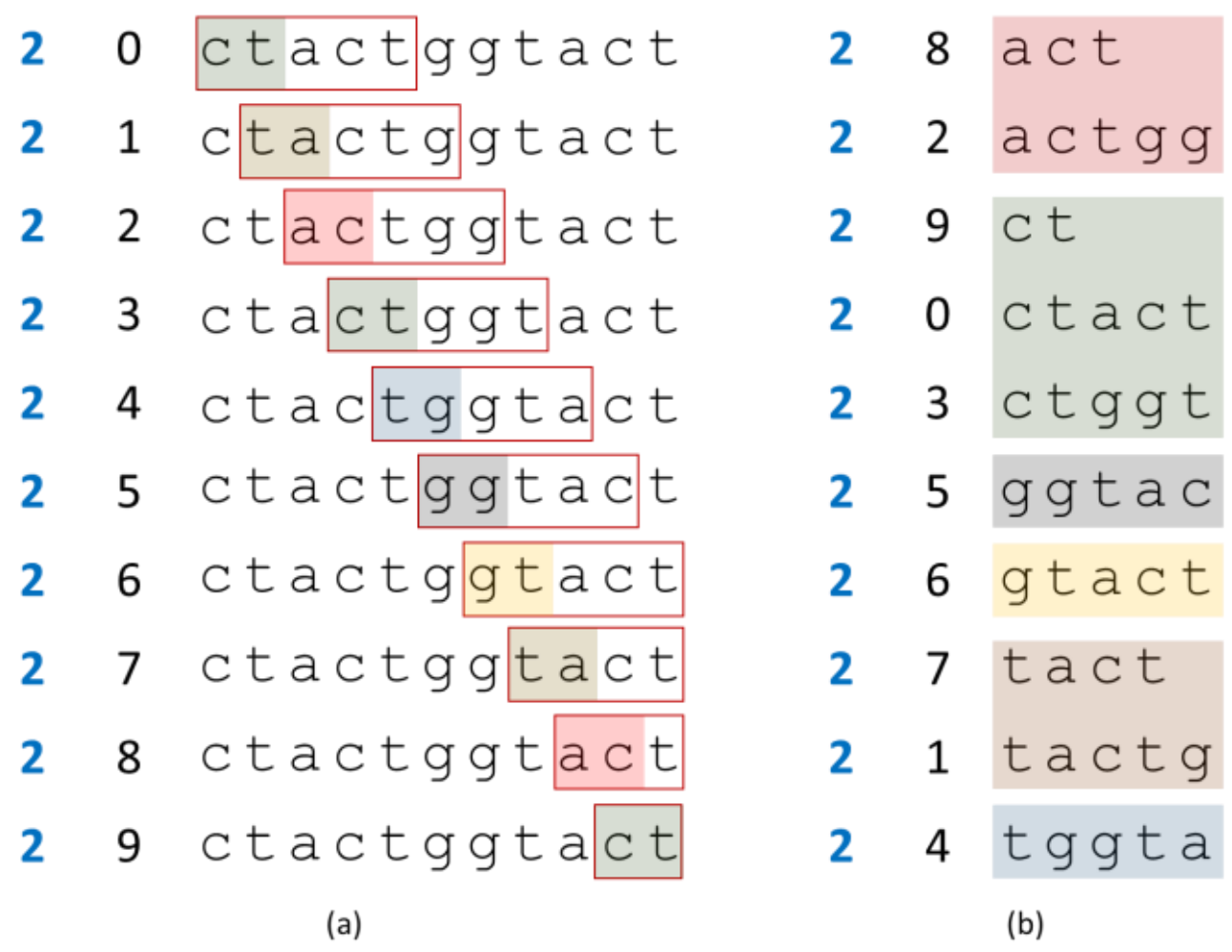

**Fig. 3** LERP-RSA construction for *ctactggtact* with Classification Level 2

structure for pattern detection and text mining purposes [27], [28]. These attributes are:

a) Classification based on the alphabet: The classification is determined by the Classification Level which is the power that the cardinality of the alphabet can be raised. For DNA sequences using the four nucleotides alphabet A, C, G and T, the classification can vary from one class, $\Sigma^{CL} = 4^0 = 1$, for classification level zero to, e.g., 16 classes, $\Sigma^{CL} = 4^2 = 16$, for classification level two with the construction of subclasses of suffix strings starting with *AA, AC, AG, AT, CA, CC, CG, CT, GA, GC, GG, GT, TA, TC, TG* and *TT*. Therefore, instead of having one class we can have 16 with significantly smaller classes, with size for each-one one sixteenth of the total, if we assume equidistribution. In the case of protein sequences, the alphabet is 20 characters based on the amino-acids encoding.

b) Network and cloud distribution based on the classes: Each class, regardless size, can be constructed or distributed independently over a local network or on the cloud. The classes can be stored and accessed when needed.

c) Full and semi parallelism: Since we have several, separate, classes the analysis and pattern detection algorithms can be executed on each class in parallel in full mode, all simultaneously, or semi-parallel mode where a block of classes is analyzed and when finished the analysis continues with the second block, etc.

d) Self-compression: When we use classification then we have in each class those suffix strings that specifically start with the class string. Therefore, the initial characters defining the class of the suffix strings in each class can be truncated and conserve space.

e) Indeterminacy: More space can be conserved for the cases that we do not care about the positions of the patterns rather than only for their existence. In this case the position indexes can be omitted.

f) The LERP-RSA can be constructed to describe multiple strings and allow the detection of patterns that exist not only in a single string but also among two or more different strings.

For many real-world cases, such as biological sequences analysis and pattern detection, it is important to perform such tasks on multiple sequences. The last attribute described above is very important for these cases since it allows to detect patterns that are not repeated per se, yet, they exist once in several sequences, making them repeated. For this purpose, we need to construct the Multivariate LERP-RSA data structure as it can be described with the following example.

Let's assume that we have two sequences *actactggtgt* and *ctactggtact* and, moreover, the LERP value is five while we have decided to use Classification Level two. In order to construct the data structure, we start with the first sequence at position zero and we use a sliding window of size five to determine the suffix strings (Fig.2.a). Additionally, for each position and suffix string we record the first two characters and we store the suffix string to the corresponding class (Fig.2.b). For example, the first five characters long substring of the first sequence is the *actac* and it will be stored in class *ac* with leading numbers to describe the sequence index (1, blue) and position in the specific sequence (0, black). We continue with the next substring *ctact* starting with *ct* which will be stored in class *ct*. We continue the process until position 9 where the substring *gt* with length two, exactly as the classification level, is the last one to be stored. The process of storing the suffix strings in each class (different colors for the example) can be performed directly or by sorting them. The same process repeats for the second sequence (Fig.3.a – Fig.3.b). Finally, the subclasses are combined together to create the lexicographically sorted Multivariate LERP-RSA (Fig.4.a) where each class is presented with different color.

## B. *ARPaD Algorithm*

After constructing the Multivariate LERP-RSA data structure we execute the All Repeated Patterns Detection algorithm. The algorithm has two versions, the recursive left-to-right and the non-recursive top-to-bottom [28]. Both versions have the same time complexity $O(n\log n)$. Since it is easier to present with an example the recursive, we will use the LERP-RSA of the previous subsection example in Fig.4.a.

First, the algorithm starts with the first class it has been created, *ac*, and counts how many strings starts with it (Fig. 4.b). Since there are four suffix strings in this class then then the class itself is a repeated pattern. The algorithm constructs a longer pattern with the first letter of the nucleotides alphabet, *a*, the *aca*. This does not exist and the algorithm continues with the other letters of the alphabet until it finds the pattern *act* which also appears four times (Fig.4.b). The process is repeated for longer patterns, starting with *acta*, until it finds the *actg* occurring twice and the longer *actgg* (Fig.4.b) which also occurs twice. With this the algorithm has discovered all repeated patterns of class *ac* or similarly starting with *ac*. The process is executed for each class and the ARPaD algorithm discovers at the end all repeated patterns (Fig.4.b). The non-recursive top-to-bottom version works in a similar way by comparing directly suffix string tuples.

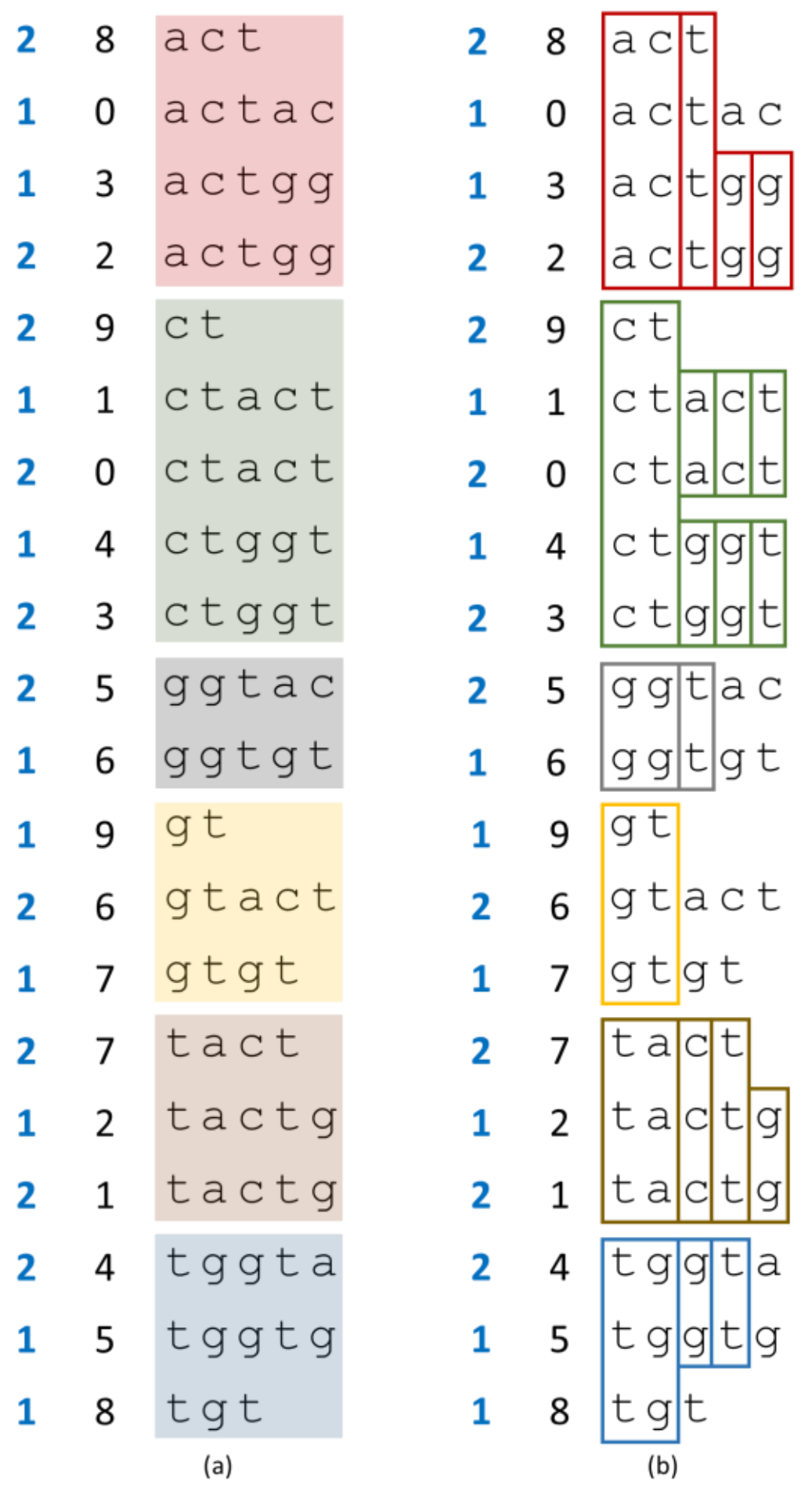

**Fig. 4** Multivariate LERP-RSA with Classification Level 2 for *actactggtgt* and *ctactggtact* (a) and ARPaD results (b)

Based on the above presented example, we can observe that ARPaD is executed on each class independently and, therefore, it can be executed in parallel. The only constrains for such execution is the available hardware, processors or cores and memory. For example, if the available resources do not allow for full parallel execution, we can start with the classes *ac* and *ta* which have the same number of suffix strings. Then we observe that class *ct* has five suffix strings while classes *gg* and *gt* have also five suffix strings combined. Therefore, we can execute in semi-parallel mode class *ct* with *gg* and when *gg* finishes, obviously before *ct*, we continue with class *gt*. This order of execution optimizes resources usage and minimizes idle time for the CPU.

Of course, we can execute ARPaD independently on each class, assuming enough resources, or even use different Classification Level per alphabet letter. This can be achieved also for datasets that significantly exceed the available local resources by using the network and/or cloud distribution. This property of LERP-RSA and ARPaD allows to use completely isolated and diversified hardware, e.g., smartphones, to analyze each class in complete isolation from other classes instead of using expensive hardware infrastructure or clustering frameworks such as Hadoop and Spark. For example, if the total size of the LERP-RSA that ARPaD should analyze is in the band of Terabytes then it is practically impossible to have such memory available for a single, parallel execution. However, if there are available 10,000 smartphones (e.g., from the students of a university) to share the LERP-RSA classes then the problem has been reduced to the band of Megabytes, which is something that smartphones can easily handle and the whole process can be executed in a single step parallel mode.

Obviously, this doesn't prevent ARPaD to be executed on a clustering framework since at each node of it specific classes can be assigned. Yet, the big difference is that LERP-RSA and ARPaD do not need any special software to control the processes execution on each node of the clustering framework or the need to constantly create and manage containers with the use of special technologies, for example, Kubernetes, Argo, Dask, etc. Simply, each node can be assigned to a specific class and return the results whenever each one has finished.

### C. SPaD Algorithm

Another important algorithm of the ARPaD family is the Single Pattern Detection (SPaD) algorithm [28]. The SPaD algorithm is mainly used for meta-analyses purposes, when we want to discover specific information in the ARPaD results or LERP-RSA, and its correctness has been proven in [28]. Moreover, especially with the LERP-RSA it can be extremely efficient with time complexity $O(1)$ with regard to the input string [28]. Although ARPaD can be executed once to detect all repeated patterns that can be stored for later meta-analyses purposes, SPaD has to be used every time we need to, e.g., check the existence of non-repeated patterns. For this purpose, we execute the SPaD directly on the LERP-RSA data structure since single occurred patterns can exist only in the LERP-RSA, if they do exist. There are two distinct cases of SPaD execution with regard to the length of the pattern we need to find; if a pattern is equal or shorter than LERP or if a pattern is longer than LERP.

Using the previously stated example we can describe the SPaD algorithm using two sample patterns with regard to their size in comparison to the LERP value. The first pattern is the *gtg*, which is not repeated pattern since we cannot find it in the ARPaD results and it is shorter than the LERP value. Since the pattern starts with *gt*, SPaD starts in the appropriate *gt* class and using the binary search algorithm approach finds the suffix string in the class, *gtact* (Fig.5.a-1). Since *gtg* is lexicographically after *gtact*, the algorithm continues in the second half of the *gt* class and finds once the pattern in the suffix string *gtgt* (Fig.5.a-2). Therefore, the pattern *gtg* exists once in the first sequence at position seven.

The next example is the *tactggtg* pattern which is longer than the LERP value. The first step is to break down the pattern under investigation to fragments of size LERP, except, of course, the last one which can be smaller. Therefore, for the particular pattern we have two fragments, the *tactg* and the *gtg*. The next step for the SPaD algorithm is to search for each fragment and record if it exists and where (Fig.5.b). If at least one of the fragments do not exist in the LERP-RSA then, obviously, the pattern does not exist in any sequence. However, if we find all fragments to occur somewhere in the LERP-RSA then SPaD has to check if the full pattern exists. In order to perform this SPaD uses the Crossed Minimax Criterion [22]. For the specific example, we can observe that the first fragment exists twice in the class *tc* and, more specifically, for the first sequence at position two and for the second sequence at position one (Fig.5.b-1). The second fragment can be found only once in class *gt* for sequence one at position seven (Fig.5.b-2). First of all, since the second fragment does not exist in the second sequence, therefore, the pattern does not exist in the second sequence. For the first sequence, the first fragment exists at position two and the second at position seven. Since the second position (7) is equal to the first position (2) plus LERP value (5), therefore, the pattern exists in the first sequence at position two.

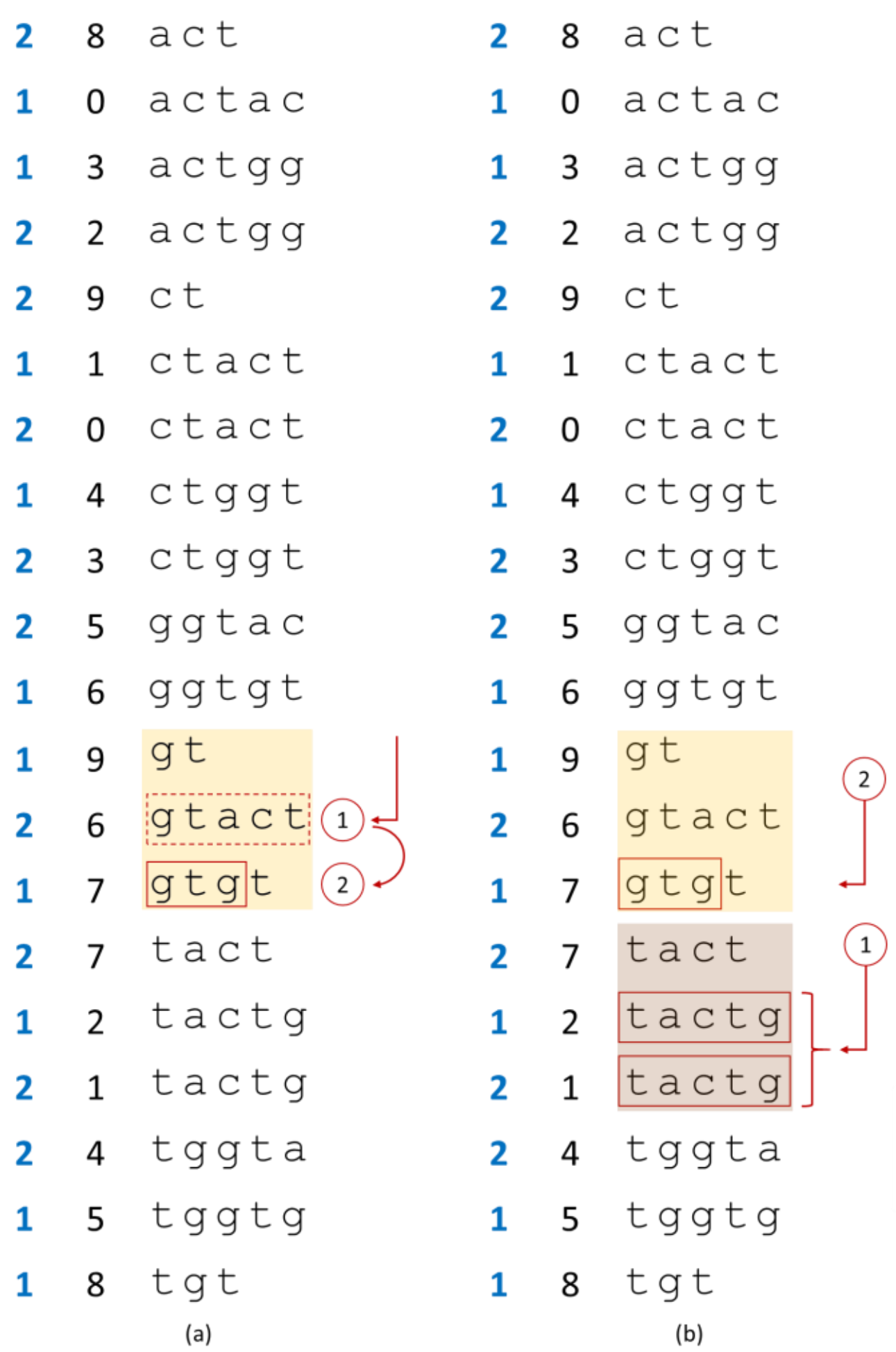

**Fig. 5** SPaD algorithm process for patterns *gtg* (a) and *tactggtg* (b)

The SPaD algorithm except of its straight forward application described above can also be used with wildcards or regular expressions, for the detection of more complex patterns. Let's assume that we want to detect all patterns with the form *t**??**tg*, where the symbol ***?*** means any character from the alphabet. Therefore, we care to find patterns such as *t**aa**tg*, *t**ac**tg*, *t**ag**tg*, *t**at**gt*, *t**ca**gt*, etc. Executing the SPaD for each combination or by using regular expressions we can detect the patterns *tactg* at positions (1, 2) and (2, 1) and *tggtg* at position (1, 5). However, when we use wildcards or we need to detect multiple patterns, the best option for optimization purposes, is the use of the MPaD algorithm of the next subsection.

### *D. MPaD Algorithm*

The Multiple Pattern Detection (MPaD) [28] algorithm is a direct extension of the SPaD. For multiple pattern detection, instead of executing in a loop the SPaD algorithm, the process is optimized with the use of the MPaD. Practically, the first step of the SPaD is extended by breaking down all patterns into fragments and adding common fragments into batches. This can help the algorithm execution because patterns can have shared fragments that they will be searched only once and if not existed a complete batch of patterns can be rejected simultaneously, instead of repeating the process. As with SPaD, MPaD can also be used with wildcards and regular expressions for more advanced pattern detection.

### *E. Meta-Analytics*

After the completion of the initial data analysis several metadata analyses can be performed. These analyses depend on several factors and the problems that we want to address such as sequence alignment, genome comparison, palindromes and tandem repeats detection, etc. The importance of the full analysis and repeated patterns detection is that it needs to be executed only once and our further, detailed, meta-analyses in the results are standalone processes. Moreover, the results can be stored on external storage media, locally or remotely on the cloud, and accessed whenever is needed, by class, without the need to repeat the analysis or access the full dataset.

In this section, three different, novel, algorithms that use as input the results of ARPaD and can be used to solve completely different problems will be introduced.

#### *1) PCR Primers Detection*

Since the scope of the algorithm presented here is to identify possible primers for PCR, it is important to search for patterns that exist in approximately the same position with a small deviation. Additionally, following the 5-prime to 3-prime PCR execution we should find patterns from both start and end of the sequence. Therefore, we use two bands that define two regions at the beginning and end of sequence, still, it can be anywhere in the full DNA sequence depending on which part of the DNA we want to amplify.

First of all, the algorithm terminates because the first for-loop runs over the finite patterns discovered by ARPaD with

```
Algorithm 1: PCR Primers Detection
Input: ARPaD Results (ARPaD-R), Primer Length
       (PL), Primer Position Bands (PPB)
Output: List of Patterns (LP)

for pattern of length PL in ARPaD-R
    if pattern exists in all sequences and
    pattern position is in PPB for all sequences then
        add pattern to LP
end for
for each pattern in LP use SPaD
    if pattern exists in another organism then
        remove pattern from LP
end for
return LP
```

a predefined length and the second over a finite filtered results list of the patterns that have specific attributes and discovered in the first loop.

The first part of the algorithm runs over the patterns of specific length. Then it checks for every pattern if the pattern exists in all sequences of the dataset and if it exists in the same position band for all sequences. If this is true then it stores the pattern to the results. When all patterns have been scanned then the second loop scans any available database with the use of SPaD to check if the pattern occurs in another organism. If the pattern exists then it is removed from the results. When all patterns have been scanned then the algorithm returns the list of results. It is important to mention that the second loop is optional since it depends on the available databases of other organisms and, moreover, since for large patterns the probability to exist in another organism and more particularly in human, is significantly small.

The algorithm is correct because in the first for-loop it checks that a pattern exists in all sequences and at the same band, regardless of exact position in the band. The reason that position bands are used and not exact positions is that because of insertion or deletion mutations the pattern can be found before or after the original position on the reference sequence. Thus, it is important to find the pattern approximate position with regard to the reference sequence. In the second for-loop the algorithm just uses the SPaD algorithm to investigate if the pattern exists in another organism (human).

The worst-case time complexity of the algorithm is $O(rs + r\log d)$ with regard to the number $r$ of patterns of length PL and occurrences $s$ in all sequences in PPB and the size $d$ of optional organism databases. It is important to mention that the factor $rs$ is many magnitudes smaller than the original size, because of the number of patterns of length PL, as it will be presented in the experimental results section. Moreover, the check of pattern position is very simple since the ARPaD results can be stored sorted by sequences index and position. Therefore, there is no need for a run over the full positions list of the pattern rather than a quick binary search for possible positions in the PPB. The second factor in the complexity is the total cost for the SPaD algorithm which is a logarithmic binary search for each pattern in LP.

#### 2) *Palindromes*

As mentioned in the previous section, CRISPR problem is related with the detection of repeated palindromes. Moreover, it is obvious that since ARPaD detects all repeated patterns the only task that needs to be performed for palindrome detection is to filter ARPaD results only for palindromes.

Algorithm 2 is simple in execution since it has a simple for-loop over every pattern found by ARPaD with length PL, and, therefore, it terminates. Inside the loop, the algorithm checks if the pattern is a palindrome. This is a trivial string problem and there are several solutions, such as using a stack, an array, etc. If the pattern is a palindrome then the algorithm verifies if it exists multiple times in the sequences of appearance and store the sequence in the results list. The list is returned when the for-loop completes. The second if-statement is optional in case we care only for palindromes that exist multiple times in every sequence.

The algorithm is correct because in every run of the loop it checks if a pattern of length PL is palindrome or not. Then it checks if it exists multiple times in every sequence. The worst-case time complexity if $O(r(p + s))$ with regard to the number of patterns $r$ and the palindrome check $p$ plus the check of multiple occurrences in every sequence.

```
Algorithm 2: Palindromes Detection
Input: ARPaD Results (ARPaD-R), Palindrome Length
        (PL), Multiple Occurrences (MO)
Output: List of Patterns (LP)

for pattern of length PL in ARPaD-R
    if pattern is palindrome then
        if pattern exists at least MO in a sequence
        then
            add pattern to LP
end for
return LP
```

```
Algorithm 3: Tandem Repeats Detection
Input: ARPaD Results (ARPaD-R), Tandem Length
        (TL), Tandem Minimum Occurrences (TMO)
Output: List of Patterns (LP)

for pattern of length TL in ARPaD-R
    for position in pattern position
        if pattern exists in TMO positions and
        pattern periodicity is TL then
            add pattern to LP
    end for
end for
return LP
```

#### 3) *Tandem Repeats*

For the detection of Tandem Repeats, Algorithm3 can be used. The algorithm runs over patterns of a specific length TL, usually very small, and checks if there are enough occurrences of at least TMO that have a periodicity of exactly the length of the tandem. In this way, longer patterns of tandems are constructed. If a pattern exists then it stored in the results list.

The algorithm terminates because it runs over the patterns set of a specific length that can be found in the ARPaD results and then over the occurrences of the pattern. The algorithm is correct because the outer loop runs over all patterns of length TL while the inner loop on the sorted occurrences of the pattern. If the for a successive TMO number of occurrences the periodicity of the pattern, i.e., occurrences, have periodicity exactly the length of the pattern then it is recorded as a tandem repeat.

The worst-case complexity is equal to a linear search over the whole dataset. If there is only a TL pattern that repeats all over the sequence then there are in total n/TL checks for the positions. Therefore, the worst-case time complexity for the algorithm is $O(ro)$ with regard to $r$ the number of patterns and $o$ the number of occurrences.

### F. *Synopsis*

The Multiple Genome Analytics Framework is based on two distinct phases. In the first phase, the first step is the construction of the Multivariate LERP-RSA data structure from the raw data of the sequences. The LERP-RSA data structure construction has a space and time complexity of $O(n\log n)$ as it has been already discussed thoroughly. In the case of the Multivariate LERP-RSA, since we have $m$ sequences of approximate length $n$, the total space complexity is $O(mn\log n)$ since the total size of the dataset, if it is considered a single sequence, is $m \times n$. However, the logarithmic part of the complexity is not equally $m \times n$ since the sequences are independent and according to Calude's theorem [24] we do not expect such long repeated patterns.

When LERP-RSA construction is completed then we execute the second step of the first phase which is the All Repeated Patterns Detection (ARPaD) algorithm. It is important to mention that both steps of the first phase are executed once during the lifecycle of the data analytics process. ARPaD has time complexity $O(n\log n)$ with regard to the dataset size and the results can be stored for any kind of meta-analytics.

Having the LERP-RSA data structure and ARPaD results stored then we can execute the second phase of the framework which is based on meta-algorithms, which can also use SPaD, MPaD algorithms, on the detected results to perform any kind of analysis such as sequence alignment, genomic comparisons, detecting primers for polymerase chain reaction process, identifying protein promoters, palindromes and tandem repeats, etc. The full execution process outline has been depicted with Fig. 6.

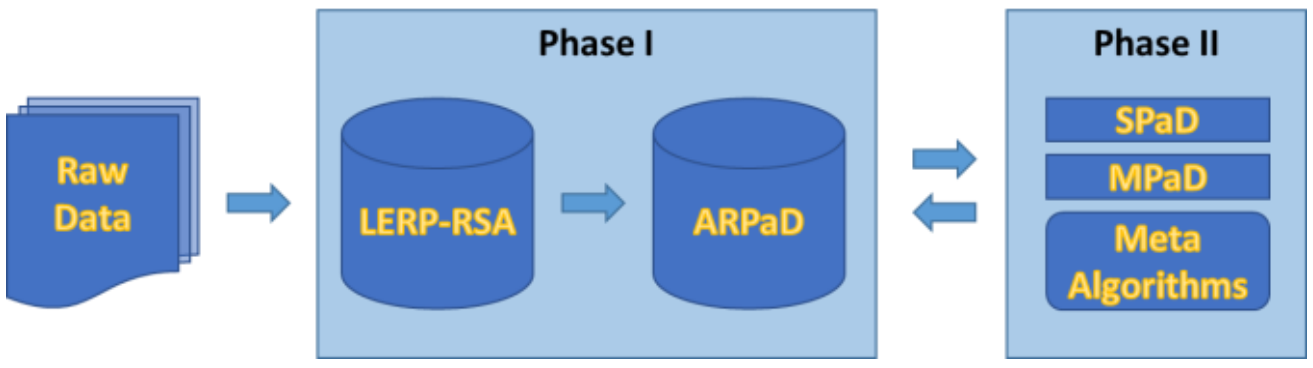

**Fig. 6** Multiple Genome Analytics Framework Workflow

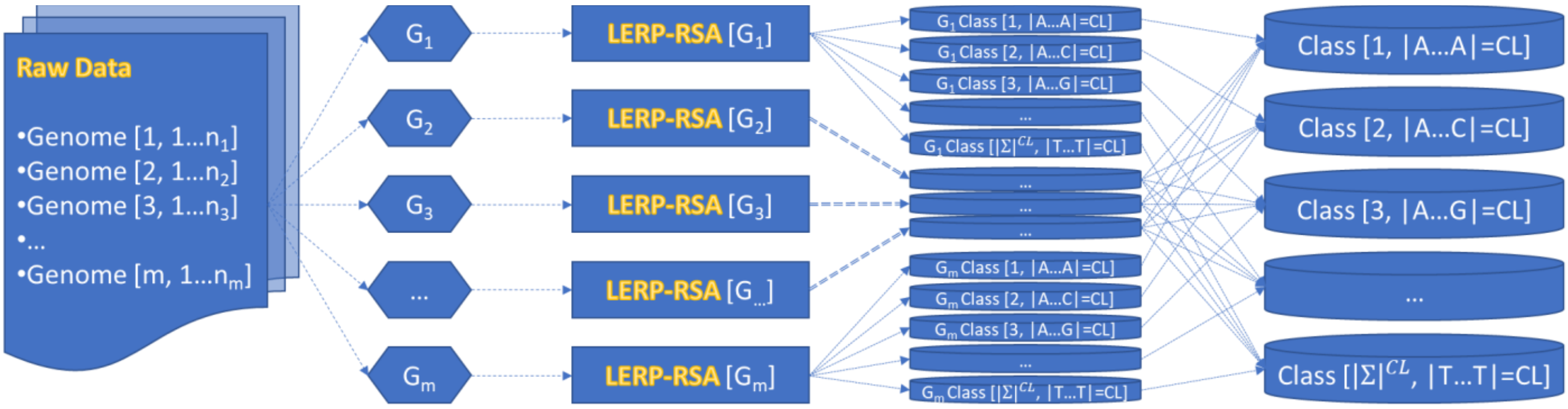

**Fig. 7** Parallel creation of LERP-RSA data structure for *m* Genome Sequences of variable length and Classification Level *CL*

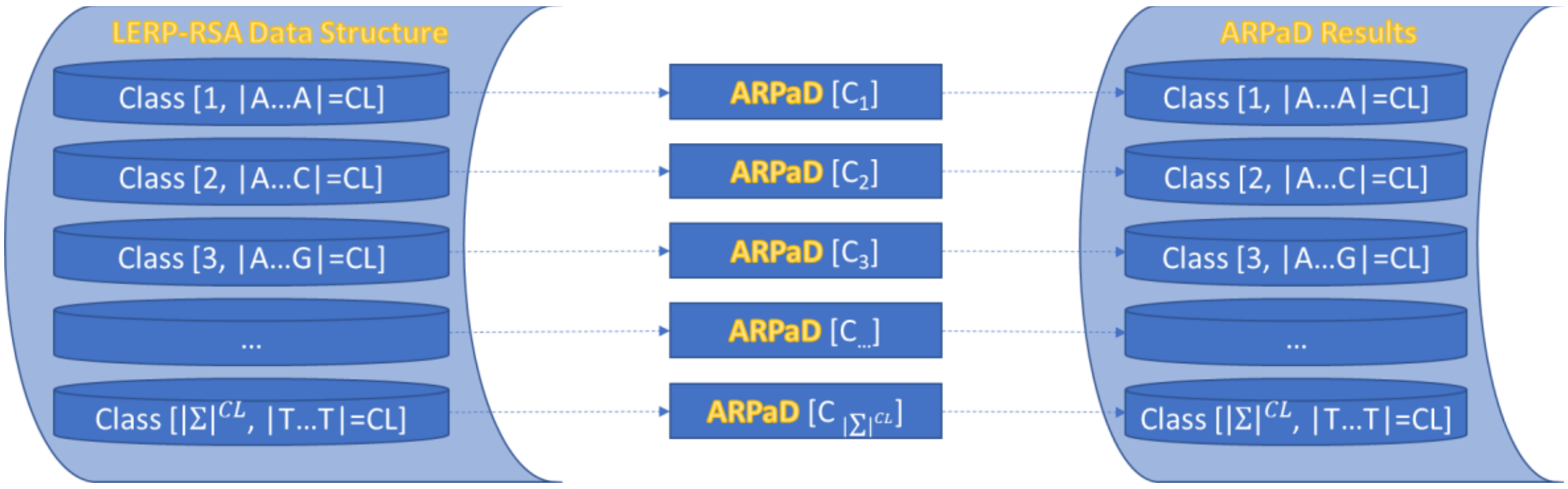

**Fig. 8** Parallel execution of ARPaD algorithm per LERP-RSA class

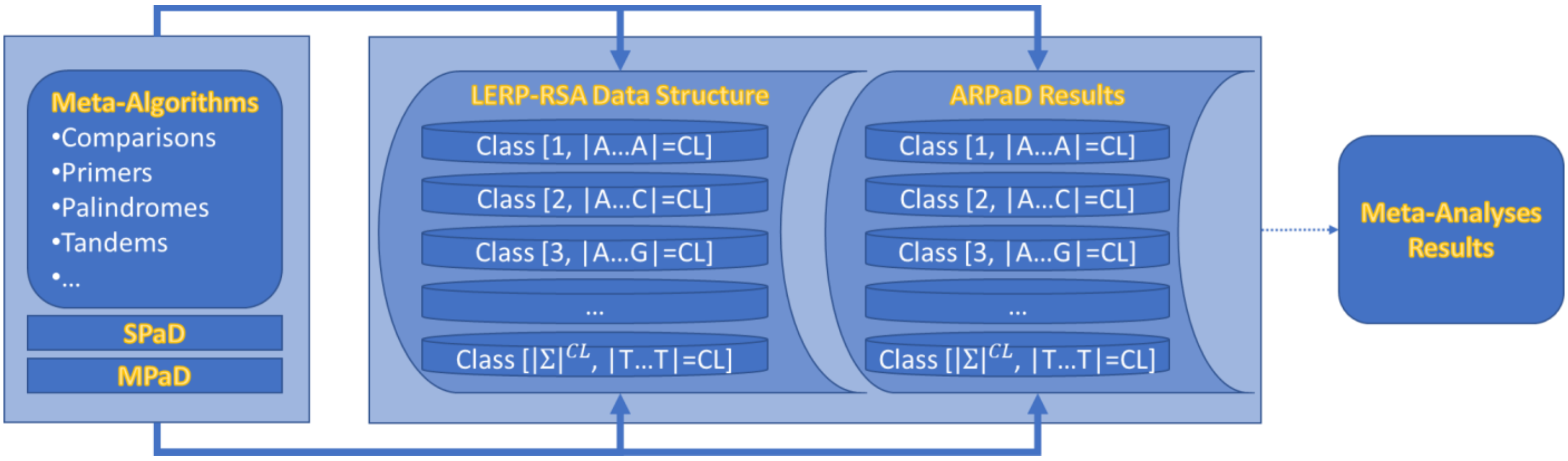

**Fig. 9** Execution of Meta-Analyses with Meta-Algorithms over the LERP-RSA and ARPaD results

In Fig. 7 the creation of the LERP-RSA data structure is represented. First the raw data is split to the $m$ genomes and then its genome is the input to the LERP-RSA creation algorithm. For each genome distinct classes are created and then the classes are combined to create the final LERP-RSA data structure. The total number of classes for each variant is $|\Sigma|^{CL}$ where the base $|\Sigma|$ is the cardinality of the alphabet, i.e., 4 for DNA/RNA sequences or 20 for proteins. It is important to mention that depending on the available hardware resources several parallelization approaches can be used. For example, LERP-RSA can be created for each genome in semi-parallel or full parallel execution depending on the number of processors or nodes in a clustering framework. Additionally, each class during the LERP-RSA creation purpose can be also be parallelized according to the available resources. ARPaD execution can also be parallelized based on available resources and number of classes, based on the Classification Level (Fig. 8).

Finally, the meta-analyses depend on several problem specific algorithms, such as Algorithms 1, 2 and 3, using LERP-RSA and ARPaD results as input and with combination of the SPaD and MPaD algorithms (Fig. 9). The whole process can also be parallelized not only for each problem but also for all problems in order to be executed simultaneously. The most important observation for the three meta-algorithms of Subsection E is that all of them use as input the ARPaD algorithm results. This is the reason for having exceptional time complexity since they use the advantage of the knowledge of all repeated patterns that have been detected in a previous phase by ARPaD.

Another, very important, observation is the difficulty to directly compare the MuGA Framework absolute time (not theoretical big-O) to other approaches. Although the theoretical time complexity it is proven to be extremely good (log-linear), the actual time cannot be compared because of the two phases of the framework and the fact that it has been built to deal with problems in a holistic approach. This means that the framework and its algorithms find all patterns that exist instead of checking if a specific pattern exists or not. For example, Algorithm 3 will find all palindromes that exist and the same is valid for Algorithm 2 for tandem repeats. Also, the first phase has to be executed only once and the results are used by all meta-algorithms concurrently, providing a very high degree of parallelization not for one algorithm, but for all algorithms that we can execute on LERP-RSA and ARPaD results to solve simultaneously many problems.

## V. Experimental Analyses and Applications

For the presentation of possible applications of the Multiple Genome Analytics Framework on different use cases, a dataset consisted from all SARS-CoV-2 (taxid 2697049) complete genome variants has been used. The dataset was recorded on May 14th, 2021, and downloaded from the National Library of Medicine at the National Center for Biotechnology Information (NCBI) [30] in its FASTA format. The dataset can be downloaded directly from the NCBI web portal by defining the appropriate parameters, i.e., (a) the virus, (b) the complete genome and (c) dataset date. However, for consistency and ease of use purposes, the NCBI constructed file can be found and downloaded directly on [34].

The recorded dataset at the specific date consists of 302,373 variants with an average variant length of 29,852 bases. However, there is one variant, the MT873050.1/USA/MA-MGH-01491/2020, which has length just 2,859 bases and it has been removed from the dataset. The total size of the dataset is approximately 9GB, three times the size of the total human genome. Although SARS-CoV-2 is a single stranded plus RNA virus, the DNA reverse transcribed sequences have been recorded in the dataset. For this reason, the standard nucleotides alphabet {A, C, G, T} has been used and the sequence strings have been cleaned from many non-standard characters such as N, R, W, etc. and replaced with a neutral symbol $ to help avoid meaningless patterns.

For the analysis, a laptop computer with an Intel i7 CPU at 2.6 GHz has been used with 16 GB RAM and an external disk of 1 TB for a semi-parallel execution, consuming approximately 40 hours. It is important though to state that the most time-consuming phase is the on-disk construction of the LERP-RSA, which took approximately 35 hours while the rest was consumed by the ARPaD. The disadvantage of using an external disk affected the performance because of the comparably low I/O transfer rate. For a broader semi-parallel execution, three computers and a tablet with approximately same configuration have been used in order to execute per computer one master class of the alphabet (A$$$, C$$$, G$$$ and T$$$) and took approximately 11 hours. The use of the semi-parallel execution significantly improved the performance, yet, it could be further improved with the use of a standard workstation computer or the use of more appropriate desktop computers. Despite that, the experiment proves that the analysis is feasible with minimal resources in acceptable time.

The Classification Level used is four, creating 256 classes because of the four characters alphabet (AAAA, AAAC, AAAG, …, TTTG, TTTT). For the specific dataset and available resources, it was not possible to use the standard value of Classification Level three, representing the 64 codon elements used for the translation process to proteins. The reason for selecting Classification Level four is to keep classes small in size and be able to analyzed by computers with 16GB RAM (or less). In an earlier initial analysis of approximately 55,000 complete variants (preprint available on bioRxiv) Classification Level three had been used, proving the versatility of the MuGA Framework and how easy is to scale up to deal with significantly larger datasets.

The results of this analysis are enormous and for practical reasons only few, interesting, use cases and meta-analyses will be presented here. The LERP value used is 60 (20 codons length). The total size of the LERP-RSA data structure on disk is 620GB, which practically means that it cannot be processed as a single class dataset. The larger class, using the predefined classification, is the TGTT with size approximately 7.2GB, which justifies the selection of Classification Level four, while the smallest is the CCGG with size approximately 210MB. Of course, the size of the data structure depends on the selection of the LERP value and can be significantly reduced based on the analyses needs and the balancing to avoid disk usage whatsoever.

A summary of the ARPaD results can be found on Table 1. There are 256 patterns with length four, as many as the classes, yet, with length five there are 1,280 instead of the expected 1,024. This happens because of the patterns which include the characters replaced with the neutral symbol $ and practically alters the alphabet size to five characters for the part of the suffix string that is longer than the Classification Level. The cumulative number of patterns with length up to

60 characters is 88.3 million approximately and the number of the total cumulative occurrences of these patterns are approximately 511.5 billion (Table 1).

Table 2 presents the most frequent 60 characters long patterns from the 64 classes of Classification Level three. The reason to present Classification Level three patterns in the table instead of the used Classification Level four is practically to keep the table as much as possible short. The patterns in the table are sorted based on the average positioning in all sequences (variants). The column next to mean positioning is the standard deviation of the pattern among all sequences, which takes values between 26 and 27 for all patterns. The next two columns are the minimum and maximum positions that the patterns have been detected in the sequences. The next column is the position that each pattern occurs in the reference sequence NC_045512.2. As we can observe, we can have some very interesting qualitative and quantitative information. For example, for the first pattern in Table 2 for class CGG, we have in total 301,269 occurrences where 151,214 happen exactly at the same position as in the reference sequence while 149,055 happen before and 1,000 after. This can help us conclude that up to the specific position most of the variants (149,055) have more deletions than insertions in the genome while the rest (1,000) have more insertions than deletions. It is also interesting that most patterns have many more mutated sequences before the average occurring position, on average 220,000, and fewer after, on average 1,050, while on average 80,000 patterns occur at the expected position.

In the same Table 2, some patterns are marked with the same color. These patterns are practically overlapping, as we can observe from their mean position which increments by one or a few more characters. These patterns can be further expanded with the use of other 60 characters long patterns or shorter patterns to form common regions in the sequences where most of the sequences are identical. Moreover, this information can be used for sequence alignment purposes, although it is a more demanding task, which will be presented in future work. The patterns in Table 2 create 12 different blocks in the SARS-CoV-2 sequence. These blocks practically separate the vast majority of the sequences to common regions and more blocks can be used with shorter patterns, something that it is expected since the patterns are from the same organism and are expected to be somehow conservable. It needs to be mentioned that this is not valid for all sequences since some may not occur in specific sequences due to mutations. Still, shorter patterns can reveal these blocks.

A very important observation regarding genome conservation of the SARS-CoV-2 organism is that the longest repeated patterns that exist among all variants at least once have length 10 bases and there are only nine (Table 3), while with length 9 they exist 61 patterns. In most cases these patterns occur in multiple positions in every variant. These can be of extreme importance since a restriction enzyme could be used to cut the SARS-CoV-2 genome at the specific patterns, degrade it and, therefore, restrict the proliferation of the virus.

Another application of the proposed methodology is the comparison of genomes among different organisms. For example, in Table 4 we have all patterns from SARS-CoV-2 that exist at least once in every variant of the virus and has length equal to 10. These patterns are compared with other organisms' genomes, using the SPaD a such as the Alpha Coronavirus (A-CoV) (taxid 693996, 1,126 total variants),

**Table 1** ARPaD Results, Patterns and Occurrences Frequencies

| L. | Patterns | Cumulative Patterns | Patterns Occurrences | Cumulative Occurrences |
|---|---|---|---|---|
| 4 | 256 | 256 | 8,979,514,845 | 8,979,514,845 |
| 5 | 1,280 | 1,536 | 8,979,360,796 | 17,958,875,641 |
| 6 | 6,377 | 7,913 | 8,979,206,429 | 26,938,082,070 |
| 7 | 28,684 | 36,597 | 8,979,051,982 | 35,917,134,052 |
| 8 | 96,118 | 132,715 | 8,978,887,799 | 44,896,021,851 |
| 9 | 220,504 | 353,219 | 8,978,696,312 | 53,874,718,163 |
| 10 | 356,768 | 709,987 | 8,978,478,185 | 62,853,196,348 |
| 11 | 459,208 | 1,169,195 | 8,978,259,310 | 71,831,455,658 |
| 12 | 531,863 | 1,701,058 | 8,978,053,216 | 80,809,508,874 |
| 13 | 591,332 | 2,292,390 | 8,977,855,568 | 89,787,364,442 |
| 14 | 645,872 | 2,938,262 | 8,977,660,436 | 98,765,024,878 |
| 15 | 698,593 | 3,636,855 | 8,977,465,502 | 107,742,490,380 |
| 16 | 750,697 | 4,387,552 | 8,977,270,080 | 116,719,760,460 |
| 17 | 802,583 | 5,190,135 | 8,977,073,859 | 125,696,834,319 |
| 18 | 854,310 | 6,044,445 | 8,976,877,424 | 134,673,711,743 |
| 19 | 905,990 | 6,950,435 | 8,976,680,835 | 143,650,392,578 |
| 20 | 957,561 | 7,907,996 | 8,976,483,481 | 152,626,876,059 |
| 21 | 1,009,096 | 8,917,092 | 8,976,285,444 | 161,603,161,503 |
| 22 | 1,060,566 | 9,977,658 | 8,976,086,825 | 170,579,248,328 |
| 23 | 1,111,944 | 11,089,602 | 8,975,887,628 | 179,555,135,956 |
| 24 | 1,163,269 | 12,252,871 | 8,975,687,751 | 188,530,823,707 |
| 25 | 1,214,552 | 13,467,423 | 8,975,487,118 | 197,506,310,825 |
| 26 | 1,265,853 | 14,733,276 | 8,975,285,952 | 206,481,596,777 |
| 27 | 1,317,114 | 16,050,390 | 8,975,084,110 | 215,456,680,887 |
| 28 | 1,368,448 | 17,418,838 | 8,974,881,778 | 224,431,562,665 |
| 29 | 1,419,819 | 18,838,657 | 8,974,678,912 | 233,406,241,577 |
| 30 | 1,471,158 | 20,309,815 | 8,974,475,528 | 242,380,717,105 |
| 31 | 1,522,507 | 21,832,322 | 8,974,271,716 | 251,354,988,821 |
| 32 | 1,573,831 | 23,406,153 | 8,974,067,684 | 260,329,056,505 |
| 33 | 1,625,123 | 25,031,276 | 8,973,863,313 | 269,302,919,818 |
| 34 | 1,676,419 | 26,707,695 | 8,973,658,918 | 278,276,578,736 |
| 35 | 1,727,777 | 28,435,472 | 8,973,455,515 | 287,250,034,251 |
| 36 | 1,779,075 | 30,214,547 | 8,973,253,405 | 296,223,287,656 |
| 37 | 1,830,382 | 32,044,929 | 8,973,057,875 | 305,196,345,531 |
| 38 | 1,881,706 | 33,926,635 | 8,972,869,112 | 314,169,214,643 |
| 39 | 1,933,040 | 35,859,675 | 8,972,683,803 | 323,141,898,446 |
| 40 | 1,984,331 | 37,844,006 | 8,972,499,669 | 332,114,398,115 |
| 41 | 2,035,636 | 39,879,642 | 8,972,316,851 | 341,086,714,966 |
| 42 | 2,086,954 | 41,966,596 | 8,972,132,927 | 350,058,847,893 |
| 43 | 2,138,316 | 44,104,912 | 8,971,948,793 | 359,030,796,686 |
| 44 | 2,189,692 | 46,294,604 | 8,971,764,688 | 368,002,561,374 |
| 45 | 2,241,072 | 48,535,676 | 8,971,579,751 | 376,974,141,125 |
| 46 | 2,292,419 | 50,828,095 | 8,971,389,770 | 385,945,530,895 |
| 47 | 2,343,742 | 53,171,837 | 8,971,193,112 | 394,916,724,007 |
| 48 | 2,395,058 | 55,566,895 | 8,970,993,383 | 403,887,717,390 |
| 49 | 2,446,380 | 58,013,275 | 8,970,790,590 | 412,858,507,980 |
| 50 | 2,497,755 | 60,511,030 | 8,970,585,542 | 421,829,093,522 |
| 51 | 2,549,149 | 63,060,179 | 8,970,377,537 | 430,799,471,059 |
| 52 | 2,600,553 | 65,660,732 | 8,970,167,730 | 439,769,638,789 |
| 53 | 2,651,922 | 68,312,654 | 8,969,956,928 | 448,739,595,717 |
| 54 | 2,703,351 | 71,016,005 | 8,969,744,785 | 457,709,340,502 |
| 55 | 2,754,827 | 73,770,832 | 8,969,531,917 | 466,678,872,419 |
| 56 | 2,806,319 | 76,577,151 | 8,969,318,628 | 475,648,191,047 |
| 57 | 2,857,829 | 79,434,980 | 8,969,104,750 | 484,617,295,797 |
| 58 | 2,909,359 | 82,344,339 | 8,968,890,580 | 493,586,186,377 |
| 59 | 2,960,959 | 85,305,298 | 8,968,676,244 | 502,554,862,621 |
| 60 | 3,012,586 | 88,317,884 | 8,968,461,663 | 511,523,324,284 |

**Table 2** Positional Descriptive Statistics for Most Frequent Patterns Per Classification Level Three Class, Length 60 and Colors for Overlapping Patterns

| I. | Class | Most Frequent Pattern with Length 60 per 3-bases Classes | Mean Pos | St.D Pos | Min Pos | Max Pos | Ref. Pos | Count | Exact | Before | After |
|---|---|---|---|---|---|---|---|---|---|---|---|
| 1 | CGG | CGGAACGTTCTGAAAAGAGCTATGAATTGCAGACACCTTTTGAAATTAAATTGGCAAAGA | 971.4 | 26 | 590 | 1027 | 989 | 301269 | 151214 | 149055 | 1000 |
| 2 | GGA | GGAAAAGTTATGTGCATGTTGTAGACGGTTGTAATTCATCAACTTGTATGATGTGTTACA | 7427.3 | 26.2 | 6495 | 7483 | 7445 | 301345 | 150809 | 149480 | 1056 |
| 3 | GGG | GGGAAATCCAACAGGTTGTAGATGCAGATAGTAAAATTGTTCAACTTAGTGAAATTAGTA | 12529.4 | 26.3 | 11451 | 12589 | 12551 | 301269 | 80808 | 219407 | 1054 |
| 4 | AAA | AAATCAGCTGGTTTTCCATTTAATAAATGGGGTAAGGCTAGACTTTATTATGATTCAATG | 14915.4 | 26.3 | 13837 | 14975 | 14937 | 301565 | 81054 | 219456 | 1055 |
| 5 | TCA | TCAGCTGGTTTTCCATTTAATAAATGGGGTAAGGCTAGACTTTATTATGATTCAATGAGT | 14918.4 | 26.3 | 13840 | 14978 | 14940 | 301438 | 80933 | 219450 | 1055 |
| 6 | CAG | CAGCTGGTTTTCCATTTAATAAATGGGGTAAGGCTAGACTTTATTATGATTCAATGAGTT | 14919.4 | 26.3 | 13841 | 14979 | 14941 | 301436 | 80933 | 219448 | 1055 |
| 7 | TCG | TCGCACCGTAGCTGGTGTCTCTATCTGTAGTACTATGACCAATAGACAGTTTCATCAAAA | 15079.4 | 26.3 | 14001 | 15139 | 15101 | 301222 | 80840 | 219330 | 1052 |
| 8 | CCG | CCGTAGCTGGTGTCTCTATCTGTAGTACTATGACCAATAGACAGTTTCATCAAAAATTAT | 15084.4 | 26.3 | 14006 | 15144 | 15106 | 301184 | 80742 | 219390 | 1052 |
| 9 | CTC | CTCTATCTGTAGTACTATGACCAATAGACAGTTTCATCAAAAATTATTGAAATCAATAGC | 15097.4 | 26.3 | 14019 | 15157 | 15119 | 301281 | 80783 | 219446 | 1052 |
| 10 | GAA | GAACTTTAAGTCAGTTCTTTATTATCAAAACAATGTTTTTATGTCTGAAGCAAAATGTTG | 15757.4 | 26.3 | 14679 | 15817 | 15779 | 301571 | 81043 | 219477 | 1051 |
| 11 | ACT | ACTTTAAGTCAGTTCTTTATTATCAAAACAATGTTTTTATGTCTGAAGCAAAATGTTGGA | 15759.4 | 26.3 | 14681 | 15819 | 15781 | 301573 | 81044 | 219478 | 1051 |
| 12 | CTT | CTTTAAGTCAGTTCTTTATTATCAAAACAATGTTTTTATGTCTGAAGCAAAATGTTGGAC | 15760.4 | 26.3 | 14682 | 15820 | 15782 | 301545 | 81036 | 219458 | 1051 |
| 13 | TAC | TACATGATGAGTTAACAGGACACATGTTAGACATGTATTCTGTTATGCTTACTAATGATA | 16089.4 | 26.3 | 15011 | 16149 | 16111 | 301673 | 81058 | 219564 | 1051 |
| 14 | AAC | AACATTAGCTGTACCCTATAATATGAGAGTTATACATTTTGGTGCTGGTTCTGATAAAGG | 20806.4 | 26.6 | 18809 | 20866 | 20828 | 301581 | 81050 | 219480 | 1051 |
| 15 | TAG | TAGCTGTACCCTATAATATGAGAGTTATACATTTTGGTGCTGGTTCTGATAAAGGAGTTG | 20811.4 | 26.6 | 18814 | 20871 | 20833 | 301566 | 81041 | 219474 | 1051 |
| 16 | AGC | AGCTGTACCCTATAATATGAGAGTTATACATTTTGGTGCTGGTTCTGATAAAGGAGTTGC | 20812.4 | 26.6 | 18815 | 20872 | 20834 | 301566 | 81040 | 219475 | 1051 |
| 17 | GCT | GCTGTACCCTATAATATGAGAGTTATACATTTTGGTGCTGGTTCTGATAAAGGAGTTGCA | 20813.4 | 26.6 | 18816 | 20873 | 20835 | 301562 | 81039 | 219472 | 1051 |
| 18 | CTG | CTGTACCCTATAATATGAGAGTTATACATTTTGGTGCTGGTTCTGATAAAGGAGTTGCAC | 20814.4 | 26.6 | 18817 | 20874 | 20836 | 301564 | 81039 | 219474 | 1051 |
| 19 | TGT | TGTACCCTATAATATGAGAGTTATACATTTTGGTGCTGGTTCTGATAAAGGAGTTGCACC | 20815.4 | 26.6 | 18818 | 20875 | 20837 | 301562 | 81038 | 219473 | 1051 |
| 20 | GTA | GTACCCTATAATATGAGAGTTATACATTTTGGTGCTGGTTCTGATAAAGGAGTTGCACCA | 20816.4 | 26.6 | 18819 | 20876 | 20838 | 301539 | 81036 | 219451 | 1052 |
| 21 | ACC | ACCCTATAATATGAGAGTTATACATTTTGGTGCTGGTTCTGATAAAGGAGTTGCACCAGG | 20818.4 | 26.6 | 18821 | 20878 | 20840 | 301549 | 81036 | 219461 | 1052 |
| 22 | CCC | CCCTATAATATGAGAGTTATACATTTTGGTGCTGGTTCTGATAAAGGAGTTGCACCAGGT | 20819.4 | 26.6 | 18822 | 20879 | 20841 | 301549 | 81035 | 219462 | 1052 |
| 23 | CCT | CCTATAATATGAGAGTTATACATTTTGGTGCTGGTTCTGATAAAGGAGTTGCACCAGGTA | 20820.4 | 26.6 | 18823 | 20880 | 20842 | 301549 | 81035 | 219462 | 1052 |
| 24 | CTA | CTATAATATGAGAGTTATACATTTTGGTGCTGGTTCTGATAAAGGAGTTGCACCAGGTAC | 20821.4 | 26.6 | 18824 | 20881 | 20843 | 301549 | 81035 | 219462 | 1052 |
| 25 | TAT | TATAATATGAGAGTTATACATTTTGGTGCTGGTTCTGATAAAGGAGTTGCACCAGGTACA | 20822.4 | 26.6 | 18825 | 20882 | 20844 | 301734 | 81061 | 219614 | 1059 |
| 26 | ATA | ATAATATGAGAGTTATACATTTTGGTGCTGGTTCTGATAAAGGAGTTGCACCAGGTACAG | 20823.4 | 26.6 | 18826 | 20883 | 20845 | 301720 | 81059 | 219602 | 1059 |
| 27 | TAA | TAATATGAGAGTTATACATTTTGGTGCTGGTTCTGATAAAGGAGTTGCACCAGGTACAGC | 20824.4 | 26.6 | 18827 | 20884 | 20846 | 301720 | 81059 | 219602 | 1059 |
| 28 | AAT | AATATGAGAGTTATACATTTTGGTGCTGGTTCTGATAAAGGAGTTGCACCAGGTACAGCT | 20825.4 | 26.6 | 18828 | 20885 | 20847 | 301600 | 81058 | 219483 | 1059 |
| 29 | ATG | ATGAGAGTTATACATTTTGGTGCTGGTTCTGATAAAGGAGTTGCACCAGGTACAGCTGTT | 20828.4 | 26.6 | 18831 | 20888 | 20850 | 301605 | 81055 | 219491 | 1059 |
| 30 | TGA | TGAGAGTTATACATTTTGGTGCTGGTTCTGATAAAGGAGTTGCACCAGGTACAGCTGTTT | 20829.4 | 26.6 | 18832 | 20889 | 20851 | 301605 | 81055 | 219491 | 1059 |
| 31 | AGA | AGAGTTATACATTTTGGTGCTGGTTCTGATAAAGGAGTTGCACCAGGTACAGCTGTTTTA | 20831.4 | 26.6 | 18834 | 20891 | 20853 | 301640 | 81079 | 219502 | 1059 |
| 32 | GAG | GAGTTATACATTTTGGTGCTGGTTCTGATAAAGGAGTTGCACCAGGTACAGCTGTTTTAA | 20832.4 | 26.6 | 18835 | 20892 | 20854 | 301639 | 81079 | 219501 | 1059 |
| 33 | AGT | AGTTATACATTTTGGTGCTGGTTCTGATAAAGGAGTTGCACCAGGTACAGCTGTTTTAAG | 20833.4 | 26.6 | 18836 | 20893 | 20855 | 301631 | 81079 | 219493 | 1059 |
| 34 | GTT | GTTATACATTTTGGTGCTGGTTCTGATAAAGGAGTTGCACCAGGTACAGCTGTTTTAAGA | 20834.4 | 26.6 | 18837 | 20894 | 20856 | 301625 | 81077 | 219489 | 1059 |
| 35 | TTA | TTATACATTTTGGTGCTGGTTCTGATAAAGGAGTTGCACCAGGTACAGCTGTTTTAAGAC | 20835.4 | 26.6 | 18838 | 20895 | 20857 | 301628 | 81079 | 219490 | 1059 |
| 36 | ACA | ACATTTTGGTGCTGGTTCTGATAAAGGAGTTGCACCAGGTACAGCTGTTTTAAGACAGTG | 20839.4 | 26.6 | 18842 | 20899 | 20861 | 301654 | 81083 | 219512 | 1059 |
| 37 | CAT | CATTTTGGTGCTGGTTCTGATAAAGGAGTTGCACCAGGTACAGCTGTTTTAAGACAGTGG | 20840.4 | 26.6 | 18843 | 20900 | 20862 | 301655 | 81084 | 219512 | 1059 |
| 38 | ATT | ATTTTGGTGCTGGTTCTGATAAAGGAGTTGCACCAGGTACAGCTGTTTTAAGACAGTGGT | 20841.4 | 26.6 | 18844 | 20901 | 20863 | 301655 | 81084 | 219512 | 1059 |
| 39 | TTT | TTTTGGTGCTGGTTCTGATAAAGGAGTTGCACCAGGTACAGCTGTTTTAAGACAGTGGTT | 20842.4 | 26.6 | 18845 | 20902 | 20864 | 301653 | 81084 | 219510 | 1059 |
| 40 | TTG | TTGGTGCTGGTTCTGATAAAGGAGTTGCACCAGGTACAGCTGTTTTAAGACAGTGGTTGC | 20844.4 | 26.6 | 18847 | 20904 | 20866 | 301645 | 81095 | 219491 | 1059 |
| 41 | TGG | TGGTGCTGGTTCTGATAAAGGAGTTGCACCAGGTACAGCTGTTTTAAGACAGTGGTTGCC | 20845.4 | 26.6 | 18848 | 20905 | 20867 | 301644 | 81094 | 219491 | 1059 |
| 42 | GGT | GGTGCTGGTTCTGATAAAGGAGTTGCACCAGGTACAGCTGTTTTAAGACAGTGGTTGCCT | 20846.4 | 26.6 | 18849 | 20906 | 20868 | 301641 | 81094 | 219489 | 1058 |
| 43 | GTG | GTGCTGGTTCTGATAAAGGAGTTGCACCAGGTACAGCTGTTTTAAGACAGTGGTTGCCTA | 20847.4 | 26.6 | 18850 | 20907 | 20869 | 301640 | 81095 | 219487 | 1058 |
| 44 | TTC | TTCTTTTGGTGGTGTCAGTGTTATAACACCAGGAACAAATACTTCTAACCAGGTTGCTGT | 23305.9 | 27 | 21311 | 23369 | 23331 | 301426 | 78794 | 221570 | 1062 |
| 45 | CGC | CGCTTGTTAAACAACTTAGCTCCAATTTTGGTGCAATTTCAAGTGTTTTAAATGATATCC | 24417.9 | 27 | 22423 | 24481 | 24443 | 301325 | 78617 | 221643 | 1065 |
| 46 | CAC | CACGTCTTGACAAAGTTGAGGCTGAAGTGCAAATTGATAGGTTGATCACAGGCAGACTTC | 24480.9 | 26.7 | 23508 | 24544 | 24506 | 301482 | 78639 | 221780 | 1063 |
| 47 | ACG | ACGTCTTGACAAAGTTGAGGCTGAAGTGCAAATTGATAGGTTGATCACAGGCAGACTTCA | 24481.9 | 26.7 | 23509 | 24545 | 24507 | 301496 | 78639 | 221794 | 1063 |
| 48 | CGT | CGTCTTGACAAAGTTGAGGCTGAAGTGCAAATTGATAGGTTGATCACAGGCAGACTTCAA | 24482.9 | 26.7 | 23510 | 24546 | 24508 | 301495 | 78638 | 221794 | 1063 |
| 49 | GTC | GTCTTGACAAAGTTGAGGCTGAAGTGCAAATTGATAGGTTGATCACAGGCAGACTTCAAA | 24483.9 | 26.7 | 23511 | 24547 | 24509 | 301499 | 78640 | 221796 | 1063 |
| 50 | TCT | TCTTGACAAAGTTGAGGCTGAAGTGCAAATTGATAGGTTGATCACAGGCAGACTTCAAAG | 24484.9 | 26.7 | 23512 | 24548 | 24510 | 301501 | 78640 | 221798 | 1063 |
| 51 | GAC | GACAAAGTTGAGGCTGAAGTGCAAATTGATAGGTTGATCACAGGCAGACTTCAAAGTTTG | 24488.9 | 26.7 | 23516 | 24552 | 24514 | 301477 | 78632 | 221779 | 1066 |
| 52 | AAG | AAGTTGAGGCTGAAGTGCAAATTGATAGGTTGATCACAGGCAGACTTCAAAGTTTGCAGA | 24492.9 | 26.7 | 23520 | 24556 | 24518 | 301492 | 78622 | 221804 | 1066 |
| 53 | GGC | GGCTGAAGTGCAAATTGATAGGTTGATCACAGGCAGACTTCAAAGTTTGCAGACATATGT | 24499.9 | 26.7 | 23527 | 24563 | 24525 | 301426 | 78639 | 221720 | 1067 |
| 54 | TGC | TGCAAATTGATAGGTTGATCACAGGCAGACTTCAAAGTTTGCAGACATATGTGACTCAAC | 24507.9 | 26.7 | 23535 | 24571 | 24533 | 301459 | 78647 | 221746 | 1066 |
| 55 | GCA | GCAAATTGATAGGTTGATCACAGGCAGACTTCAAAGTTTGCAGACATATGTGACTCAACA | 24508.9 | 26.7 | 23536 | 24572 | 24534 | 301448 | 78645 | 221737 | 1066 |
| 56 | CAA | CAAATTGATAGGTTGATCACAGGCAGACTTCAAAGTTTGCAGACATATGTGACTCAACAA | 24509.9 | 26.7 | 23537 | 24573 | 24535 | 301506 | 78665 | 221772 | 1069 |
| 57 | GAT | GATAGGTTGATCACAGGCAGACTTCAAAGTTTGCAGACATATGTGACTCAACAATTAATT | 24515.9 | 26.7 | 23543 | 24579 | 24541 | 301497 | 78655 | 221776 | 1066 |
| 58 | AGG | AGGTTGATCACAGGCAGACTTCAAAGTTTGCAGACATATGTGACTCAACAATTAATTAGA | 24518.9 | 26.7 | 23546 | 24582 | 24544 | 301469 | 78640 | 221763 | 1066 |
| 59 | GCC | GCCATCCTTACTGCGCTTCGATTGTGTGCGTACTGCTGCAATATTGTTAACGTGAGTCTT | 26311.8 | 27.1 | 24317 | 26375 | 26337 | 301498 | 78292 | 222110 | 1096 |
| 60 | CCA | CCATCCTTACTGCGCTTCGATTGTGTGCGTACTGCTGCAATATTGTTAACGTGAGTCTTG | 26312.8 | 27.1 | 24318 | 26376 | 26338 | 301452 | 78288 | 222068 | 1096 |
| 61 | ATC | ATCCTTACTGCGCTTCGATTGTGTGCGTACTGCTGCAATATTGTTAACGTGAGTCTTGTA | 26314.8 | 27.1 | 24320 | 26378 | 26340 | 301505 | 78308 | 222099 | 1098 |
| 62 | TCC | TCCTTACTGCGCTTCGATTGTGTGCGTACTGCTGCAATATTGTTAACGTGAGTCTTGTAA | 26315.8 | 27.1 | 24321 | 26379 | 26341 | 301504 | 78308 | 222098 | 1098 |
| 63 | GCG | GCGCTTCGATTGTGTGCGTACTGCTGCAATATTGTTAACGTGAGTCTTGTAAAACCTTCT | 26323.8 | 27.1 | 24329 | 26387 | 26349 | 301282 | 78216 | 221968 | 1098 |
| 64 | CGA | CGATTGTGTGCGTACTGCTGCAATATTGTTAACGTGAGTCTTGTAAAACCTTCTTTTTAC | 26329.8 | 27.1 | 24335 | 26393 | 26355 | 301184 | 78221 | 221865 | 1098 |

Alpha Influenza virus (taxid 197911, 9,763 total variants), Human Orthopneumovirus (HRSV) (taxid 11250, 2,490 total variants), MERS Coronavirus (taxid 1335626, 591 total variants) [30] and the human genome (GRCh38.p12) [29]. In order to check if the patterns exist in other organisms, the SPaD and MPaD algorithms have been used, having as input the specific patterns. As we can observe at Table 4, Alpha Coronavirus has all patterns in common with SARS-CoV-2, except one. Four of them occur in very few variants of A-CoV while the rest in many more. Alpha Influenza although has six common patterns with SARS-CoV-2, they occur in very few variants of the virus. Human Orthopneumovirus has only two common patterns again with few occurrences. Interesting is the case of the MERS coronavirus that has five common patterns with SARS-CoV-2, where two of them occur in almost all MERS-CoV variants and the other three only in one or two variants. What it looks impressive is that all patterns exist in the human genome too, with different number of occurrences varying from 467 up to 14,325. Nonetheless, this observation could be probabilistically expected because of the very short pattern length and the significantly large size of the human genome. Still, slightly longer common patterns cannot be found in both the virus and the human genome.

Possible application of this information is the determination of primers for PCR analyses. Since the patterns exist in all SARS-CoV-2 variants they can be used in pairs to amplify the largest part of the virus. However, if used with human DNA sample then PCR is not possible since human genome could also be amplified. This can be bypassed with the use of longer patterns, e.g., with length 60 as the pattern in Table 2, that do not exist in the human genome. Yet, since these patterns are not present in all SARS-CoV-2 variants, two couples must be used that cover all possible cases. This can help to use PCR not just on specific SARS-CoV-2 proteins but on much larger parts of the genome. For example, if we use Algorithm 1 with the Primer Length parameter equal to the shorter, but lengthy enough, 30 characters long patterns, then the GTGCTGGTAGTACATTTATTAGTGATGAAG and GCGTGTAGCAGGTGACTCAGGTTTTGCTGC patterns will be found in the results, occurring approximately at positions 934 and 27,039 respectively, which are capable to amplify approximately 87% of the genome.

Another category of interesting patterns are the palindromes. In total they have been detected 638 repeated palindromes with length from 12 bases up to 30, including trivial palindromes, for example, continuous A or palindromes of type AA…AA$AA…AA, where $ can be any character. No test for palindromes of more than 30 or less than 12 letters has been executed because typically they form trivial and not important palindromes. In Table 5, a list of the most frequent, non-trivial, palindromes are presented. These indicative palindromes are very easy to be extracted from the ARPaD results using, for example, Algorithm 2 or query executors with regular expressions. Additionally, as it can be observed from Table 1, there are approximately 19 million patterns that need to be checked with length from 12 up to 30, which means that the required checks are 1,000 fold less than the total size of the dataset. This is valid also in the case of applying the method on a single sequence. Of course, there are some infrequent palindromes (Table 5) that are repeated just twice in the full dataset and such palindromes are in total 262.

**Table 3** Longest Patterns Existing in Every Variant of SARS-CoV-2

| Longest Patterns with Appearance at least once in Every SARS-CoV-2 Variant | Total Pattern Occurrences |
|---|---|
| ATGCTGTTGT | 905,838 |
| ATGGTAATGC | 906,103 |
| GAAGAAGCTA | 602,515 |
| TAAACGAACT | 1,061,910 |
| TATGGTGCTA | 603,948 |
| TCAACTCAGG | 901,394 |
| TGGACAACAG | 1,202,518 |
| TGGTGTTTAT | 1,207,566 |
| TTTTATGTCT | 604,441 |

**Table 4** Comparison of Longest Patterns among different Organisms

| Longest Patterns with Appearance at least once in Every SARS-CoV-2 Variant | Organism Genome | | | | |
|---|---|---|---|---|---|
| | A-CoV \|1,126\| | Alpha Influenza \|9,763\| | HRSV \|2,490\| | MERS-CoV \|591\| | GRCh38.p12 \|1\| |
| ATGCTGTTGT | 169 | 4 | 16 | 0 | 5,080 |
| ATGGTAATGC | 774 | 6 | 0 | 584 | 5,640 |
| GAAGAAGCTA | 52 | 52 | 160 | 2 | 3,813 |
| TAAACGAACT | 6 | 0 | 0 | 0 | 467 |
| TATGGTGCTA | 760 | 0 | 0 | 0 | 1,932 |
| TCAACTCAGG | 0 | 0 | 0 | 1 | 3,184 |
| TGGACAACAG | 5 | 13 | 0 | 0 | 7,302 |
| TGGTGTTTAT | 383 | 3 | 0 | 588 | 5,654 |
| TTTTATGTCT | 22 | 34 | 0 | 2 | 14,325 |

**Table 5** Indicative Most and Least Frequent Palindromes of SARS-CoV-2

| Length | Occurrences | Pattern |
|---|---|---|
| 12 | 302,210 | GTGTTAATTGTG |
| 13 | 302,089 | CAAACTGTCAAAC |
| 12 | 302,071 | AGATTGGTTAGA |
| 13 | 301,930 | TTTTGGTGGTTTT |
| 15 | 301,917 | ATTTTGGTGGTTTTA |
| 12 | 301,897 | TCAATGGTAACT |
| 12 | 301,886 | AATATCCTATAA |
| 12 | 301,770 | ATAAACCAAATA |
| 13 | 301,662 | AATTTGTGTTTAA |
| 15 | 301,655 | GAATTTGTGTTTAAG |
| 13 | 301,615 | TTGAAGAGAAGTT |
| 12 | 301,564 | AAAACAACAAAA |
| 12 | 301,552 | AGTGTAATGTGA |
| 12 | 301,547 | TTCAGTTGACTT |
| 12 | 301,447 | ATGACTTCAGTA |
| 14 | 301,438 | AATGACTTCAGTAA |
| 13 | 301,291 | AAAGACACAGAAA |
| 12 | 301,248 | CAAGAAAAGAAC |
| 13 | 300,778 | AAATCACACTAAA |
| 16 | 300,716 | CAATGACTTCAGTAAC |
| 13 | 300,702 | AGACGACAGCAGA |
| 18 | 300,694 | TCAATGACTTCAGTAACT |
| 20 | 300,147 | CTCAATGACTTCAGTAACTC |
| 12 | 299,248 | GACTCAACTCAG |
| 13 | 298,917 | TTCTAACAATCTT |
| 13 | 298,603 | TAACTTCTTCAAT |
| 13 | 294,659 | TCAGACTCAGACT |
| 15 | 294,586 | ATCAGACTCAGACTA |
| 12 | 2 | TGGACAACAGGT |
| 13 | 2 | ATACCAGACCATA |
| 13 | 2 | GTAGTGGGTGATG |
| 13 | 2 | TTGAAGCGAAGTT |
| 14 | 2 | AGCAAGTTGAACGA |
| 15 | 2 | CTTGAAGAGAAGTTC |
| 16 | 2 | AAGCAAGTTGAACGAA |
| 17 | 2 | TATCAGACTCAGACTAT |
| 18 | 2 | AAAGCAAGTTGAACGAAA |
| 19 | 2 | AATATCATCCCTACTATAA |

**Table 6** Indicative Tandem Repeats of SARS-CoV-2 per Length and Type

| Type | Length | Occurrences | Pattern |
|---|---|---|---|
| 3x3 | 9 | 301,943 | CTGCTGCTG |
| 3x3 | 9 | 301,236 | AGTAGTAGT |
| 3x3 | 9 | 300,970 | CGACGACGA |
| 3x3 | 9 | 300,920 | GACGACGAC |
| 3x3 | 9 | 13 | TAGTAGTAG |
| 4x3 | 12 | 126 | ACAAACAAACAA |
| 3x4 | 12 | 86 | GTTGTTGTTGTT |
| 3x4 | 12 | 83 | CTTCTTCTTCTT |
| 3x4 | 12 | 80 | TAATAATAATAA |
| 3x4 | 12 | 23 | GAAGAAGAAGAA |
| 4x3 | 12 | 12 | AACCAACCAACC |
| 3x4 | 12 | 8 | AGAAGAAGAAGA |
| 4x3 | 12 | 2 | AGAAAGAAAGAA |
| 3x5 | 15 | 20 | CAACAACAACAACAA |
| 4x4 | 16 | 62 | CAAACAAACAAACAAA |

Finally, in Table 6 some examples of tandem repeats are presented, as they have been detected using Algorithm 3. As with palindromes, the tandem repeats have been identified as repeated patterns and it is very easy to be filtered from the ARPaD results. Tandem repeats of length from 9 characters up to 25 have been spotted with types of repeats, such as, 3 characters by 3 times, 3x4, 4x3, 3x5, 5x3, 4x4, 3x6, 6x3, 4x5, 5x4 and 5x5. As we can observe in Table 6 from the examples, some tandem repeats are very frequent, practically they occur in most variants or multiple times per variant, while there are some others that are extremely rare. Although it is mentioned that it is possible to detect tandem repeats as repeated patterns, the obvious argument is what will happen if a tandem repeat is not repeated. Such cases are also easy to be detected because if, for example, a tandem repeat of three characters by six (or more repetitions) exist, with total length 18 characters, then its sub-patterns of three by three (or more repetitions) of nine characters have been detected as repeated patterns and since one occurs exactly after the end of the other one then they form a longer pattern and, therefore, tandem repeat.

## VI. Conclusions

The current paper presents the Multiple Genome Analytics Framework, which is a combination of data structures and algorithms specifically created for advanced text mining and pattern detection in discrete sequences that are adapted for biological sequences. More particularly, the purpose of the paper is to present a framework that can be used as the foundation of solving many different string problems in bioinformatics concurrently. As an example of possible uses, the analysis of more than 300,000 variants of the complete SARS-CoV-2 genome has been used. Using ordinary computers, it has been presented that it is possible to perform advanced pattern detection and produce results that can be fed as input to meta-algorithms or used indirectly from other methodologies to perform even more detailed or diverse meta-analyses.

The main innovation of the proposed framework is the divide and conquer approach with the use of the special LERP-RSA data structure and ARPaD algorithm. Although the construction of the LERP-RSA, on disk, could be relatively slow for commodities computers, yet, the advantage is the fact that it can be constructed off-line and stored on disk to be used for future purposes. Nevertheless, with the use of larger Classification Level with more classes of smaller sizes, the use of disk can be totally omitted and the full process can be executed directly in memory and possibly in parallel, depending on available resources. Additionally, it has been presented that with the use of LERP-RSA data structure and the single execution of ARPaD algorithm all repeated patterns can be detected, forming a database of results that meta-algorithms, with the help of SPaD and MPaD algorithms, can filter and explore to perform several meta-analyses. Both LERP-RSA data structure and ARPaD algorithm are very efficient and can produce the results in a few hours using commodity hardware while the meta-algorithms can perform various analyses in a few seconds or minutes because of the advantage of the repeated patterns knowledge.

It has been proven that the framework introduced here can address the requirements of the problem as defined in Section III. First of all, it can be executed on commodity computers with limited resources and, therefore, keep the cost low. Moreover, it can scale up to deal with larger datasets, either by simply changing initial parameters such as the LERP value and Classification Level or with insignificant cost in hardware. Finally, and more importantly, it can address several different problems concurrently by utilizing and optimizing the available hardware.

The scope of the current work is to unveil the potential benefits from the use of the Multiple Genome Analytics Framework for bioinformatics and computational biology purposes in order to solve string problems and identify hidden patterns. In future work a more detailed and thorough description of usage on additional problems will be presented with more custom-made methodologies and algorithmic variations.

## VII. Declarations

Funding: No funds, grants, or other support was received.

Conflicts of interest/Competing interests: The author has no relevant financial or non-financial interests to disclose.

Availability of data and material: The datasets generated during and/or analyzed during the current study are available in the NCBI repository [29], [30]. The downloaded SARS-CoV-2 dataset can be found additionally on Kaggle [34].